# A native chemical chaperone in the human eye lens

Eugene Serebryany[1], Sourav Chowdhury[1], Nicki E. Watson[2], Arthur McClelland[2], and Eugene I. Shakhnovich[1] *

[1]Department of Chemistry and Chemical Biology, Harvard University, Cambridge, MA, USA
[2]Center for Nanoscale Systems, Harvard University, Cambridge, MA, USA
* To whom correspondence should be addressed: shakhnovich@chemistry.harvard.edu

**Abstract**

Cataract is one of the most prevalent protein aggregation disorders and still the biggest cause of vision loss worldwide. The human lens, in its core region, lacks turnover of any cells or cellular components; it has therefore evolved remarkable mechanisms for resisting protein aggregation for a lifetime. We now report that one such mechanism relies on an unusually abundant metabolite, *myo*-inositol, to suppress light-scattering aggregation of lens proteins. We quantified aggregation suppression by *in vitro* turbidimetry and characterized both macroscopic and microscopic mechanisms of *myo*-inositol action using negative-stain electron microscopy, differential scanning fluorometry, and a thermal scanning Raman spectroscopy apparatus. Given recent metabolomic evidence that it is dramatically depleted in human cataractous lenses compared to age-matched controls, we suggest that maintaining or restoring healthy levels of *myo*-inositol in the lens may be a simple, safe, and widely available strategy for reducing the global burden of cataract.

**Introduction**

Cataract disease afflicts tens of millions of people per year and is the worldwide leading cause of blindness.[1] The disease is caused by light-scattering aggregation of the extremely long-lived crystallin proteins in the lens.[2] This aggregation is associated with accumulation of post-translational modifications, including disulfide formation, Trp oxidation, deamidation, Asp isomerization, Lys derivatization, truncation, and various others.[3-5] Most age-onset cataract forms in the oldest, nuclear (or core) region of the lens.[2] This region lacks all organelles, as well as protein synthesis and degradation capacity, and its proteome is composed ~90% of crystallins.[6] The most thermodynamically stable in humans is γD-crystallin (HγD), and it is particularly abundant in the nuclear region of the lens and enriched in insoluble aggregates.[3] HγD consists of two homologous intercalated double-Greek key domains. The N-terminal domain is less stable and derives part of its stability from the hydrophobic domain interface.[5,7]

In prior research, we have developed a physiologically relevant *in vitro* model of cataract-associated aggregation and investigated in molecular mechanistic detail the unfolding pathway and probable interactions involved.[8-11] Variants of HγD that cause congenital cataracts cluster near the N-terminal β-hairpin.[5] These variants include W42R.[12] The vast majority of cataracts are not congenital; they arise from age-related changes in the lens, including Trp oxidation.[4] We have shown that the oxidation-mimicking W42Q variant and the congenital W42R produce similarly strong aggregation under physiologically relevant oxidizing conditions.[10] We have shown that this aggregation crucially depends on formation of a non-native internal disulfide bond, which locks a partially unfolded aggregation precursor conformation,[10] and, furthermore, due to latent oxidoreductase activity in this and likely other crystallins, this conformational transition depends on the redox state of the crystallin proteome.[11]

Despite intense research interest,[13-15] no preventative or therapeutic drugs against cataract have been approved to-date, leaving surgery as the only treatment option. Aggregate costs of cataract surgery are high in developed countries,[16] while availability and outcomes are often poor in developing ones.[1,17] Hence, most cataract around the world remains untreated.[18] Delaying the age of onset of the disease by ~14% would cut the need for surgery by half by pushing the onset of blindness beyond the average lifespan.[18]

However, the search for prophylactic treatments has been hampered by three major challenges. Any such treatment must be (1) able to permeate into the nuclear region of the lens where most cataract happens; (2) simple to use (e.g., eye drops) and extremely safe; and (3) stable and cheap, to ensure it is widely available. The lens is a tightly packed tissue, and a diffusion barrier that forms around middle age prevents almost all externally applied compounds from penetrating into the lens nucleus.[19] Unless a drug lies within the highly constrained space of lens-permeating compounds, it is unlikely to be globally useful.

We therefore wondered whether the small molecule metabolome of the lens cytoplasm itself has evolved to produce or concentrate metabolites that suppress crystallin aggregation. Strong evolutionary pressure to preserve clear vision has produced multiple mechanisms to delay crystallin aggregation in the lens for as long as possible. Prior research focused largely on the crystallins themselves: the thermodynamic stability, kinetic stability, and redox activity of monomeric γ-crystallins;[5,20] the stabilizing oligomerization and destabilizing deamidation of β-crystallins;[21] and the passive chaperone role of α-crystallins.[22] Aggregation of crystallins, like that of many other proteins, can be modulated by osmolytes or other small metabolites.[23,24] Metabolomic analyses showed certain small metabolites to be unusually abundant in the lens, notably *myo*-inositol[25,26] – which, however, is greatly depleted in cataractous lenses.[26,27] We now report that *myo*-inositol acts as a chemical chaperone, delaying and slowing aggregation of cataract-associated human γD-crystallin variants. At the microscopic level, it does this by making the protein's more vulnerable N-terminal domain more resistant to localized thermal misfolding. Macroscopically, it does not alter aggregate morphology, but reduces either the number or the size of light-scattering particles, depending on concentration. *Myo*-inositol is one of the very few molecules known to permeate into the lens from the outside, by both passive and active transport.[28,29] It therefore has the potential to become a safe, simple, and widely available prophylactic against cataract.

**Results**

**Small carbohydrates suppress γD-crystallin aggregation**

Initial observations with glucose led us to suspect that small sugars and their metabolites, even at low-millimolar concentrations, could meaningfully suppress crystallin aggregation in the lens, thus delaying the onset of cataract. We used W42Q HγD as the *in vitro* model of cataract-associated aggregation, starting from a well-folded native state and at physiological pH and temperature, to compare the effects of a variety of small carbohydrates on crystallin aggregation. **Figure 1A** shows a representative set of normalized turbidity traces demonstrating strong and dose-dependent suppression of turbidity development by *myo*-inositol. As shown in **Figure 1B**, even highly similar compounds, at 100 mM each, had widely varying effects, from strong suppression to moderate enhancement of aggregation. Solution turbidity at the end of the 4.5 h incubation was measured in a plate reader (BioTek), and the turbid solution was then clarified by centrifugation. Solution turbidity is a function of both the amount of aggregated protein and the morphology and size of the aggregates. However, the amount of residual protein in the clarified supernatant had a strong linear inverse correlation with the solution turbidity, suggesting these compounds acted by inhibiting the aggregation process, not changing aggregate morphology.

That the ability to suppress crystallin aggregation varied among chemically similar compounds strongly suggests specific protein-metabolite interactions. Thus, *myo*-inositol was much more effective than mannitol – a sugar alcohol with the same number of carbons and hydroxyls, but a linear rather than cyclic structure. Galactose was one of the strongest aggregation suppressors, yet its derivative IPTG enhanced aggregation.

A more detailed comparison of dose-response curves for *myo*-inositol compared to glycerol is shown in **Figure 1C**. Tsentalovich *et al.* reported 24.6 ± 6.7 µmol of *myo*-inositol per gram of wet lens in the nuclear region, dropping to just 1.9 ± 1.8 µmol/g in the nuclear region of cataractous lenses.[26] Assuming a water content of

~60% and a total of ~50% of the water in a bound state, per prior literature,[30] we arrive at an estimate of 82 ± 22 µM *myo*-inositol in the free water fraction of the healthy human lens. In this concentration range, *myo*-inositol already had a very significant effect in our *in vitro* assay, suppressing the rate of turbidity development by ~30%. Additional metabolites present in the lens at lower levels, including *scyllo*-inositol and glucose, likely increase the total effective concentration range of aggregation suppressor molecules. Glycerol, which previous analyses have shown to also be abundant in the lens,[25] was much less effective. *Scyllo*-inositol offered no advantage over *myo*-inositol (**Figure S1**) and was not pursued further owing to its lower solubility and higher cost. Note that we have previously shown the existence of a master curve that links the lag time, rate, and end-point extent of aggregation.[31]

**Aggregation suppression is robust to details of the assay**

We next checked whether aggregation suppression was specific to the W42Q HγD variant or more broadly effective. Dose-dependent suppression of oxidative aggregation was observed in a variety of aggregation-prone HγD constructs. We have previously demonstrated that, while the W42Q variant aggregates on its own at sufficient concentrations, the wild-type protein, which does not aggregate, can catalyze W42Q aggregation.[9,11] **Figure 1D** shows that low [W42Q] aggregating in the presence of high [WT] is also suppressed by *myo*-inositol. The W130Q variant, which mimics oxidation of the cognate tryptophan in the C-terminal domain, was likewise rescued. Moreover, oxidative aggregation of the isolated, wild-type N-terminal domain of HγD was suppressed to approximately the same extent. Of these three, the W42Q + WT mixture showed slightly less suppression at lower [*myo*-inositol], perhaps attributable to the excess WT serving as a decoy for the protein-metabolite interaction. Since the N-terminal domain of HγD derives part of its stability from its interface with the homologous C-terminal domain, and aggregation involves a non-native disulfide between conserved Cys residues in the N-terminal domain, we postulate that diverse core variants converge on similar aggregation precursor conformations and differ mostly in the activation energy required to escape the native state and populate this intermediate. Since the two Cys residues forming the aggregation-determining disulfide (Cys32 and Cys41) are conserved across the γ-crystallins, we expect that the aggregation pathway is general for this protein family and predict that *myo*-inositol and its isomers and derivatives suppress aggregation of these abundant lens-specific proteins. This generality implies that aggregation due to a wide variety of cumulative covalent modifications to the wild-type protein in the most common types of age-related cataract should likewise be suppressed.

**Macroscopic mechanism of aggregation suppression**

We used negative-stain transmission electron microscopy to investigate the effect of *myo*-inositol on the aggregation process. Notably, a small number of large condensed aggregate particles were present even in the control sample (**Figure 2A**); these particles form during storage and incubation in the absence of reducing agent and contribute very little to overall turbidity. Samples from the turbid solutions (**Figure 2B,C**) showed much greater texturing on the survey micrographs and are replete with smaller, less mature aggregates numerous enough to scatter light. Images in Figure 2D-G give a sense of the likely assembly process of the aggregates.

We next carried out morphometry of the aggregates, using a total of 103 separate images from duplicate aggregated samples with 0, 100, or 250 mM *myo*-inositol. A double-blind was set up to minimize human bias: the microscopist did not know which sample came from which treatment condition, and the image analyst did not know which images came from which sample. Aggregates were visually classified into globular/collapsed or fibrillar/extended based on whether a curve could be clearly traced from one end of a fibril to the other; aggregates composed of multiple fibrils or featuring intra-chain interactions that obscured one or both ends were considered globular. For this reason, the fibrillar aggregates tended to be short in length.

We present two alternative quantification approaches: statistics of the size and number of aggregates of either type by image (**Figure 3A,C**) and the size distributions overall with particles ranked from largest to smallest (**Figure 3B,D** and **Table S2**). *Myo*-inositol produced no qualitative change in the morphology of the aggregates. It did, however, alter the size distributions in more subtle but mechanistically telling ways. At the intermediate, near-physiological concentration (100 mM), the fibrillar aggregates became shorter and the globular fewer in number. By contrast, globular aggregates in the intermediate range (rank ~25-55 in **Figure 3B**) grew slightly larger than without the inhibitor. At the high *myo*-inositol concentration (250 mM), on the other hand, the fibrillar aggregates actually lengthened slightly compared to the no-inositol control (**Figure 3D**), but the larger, globular aggregates shrank across the board (**Figure 3B**) and covered significantly less total grid area per image (**Figure 3A**).

The shapes of the size distributions for the globular aggregates under the three treatment conditions (**Figure 3B**) were compared by two-sided Kholmogorov-Smirnov tests. These tests indicated statistically likely differences between 0 and 100 mM *myo*-inositol (p = 0.0032) and between 100 and 250 mM *myo*-inositol (p = 0.0048). The shapes of the aggregate size distributions at 0 and 250 mM *myo*-inositol were not statistically distinct (p = 0.22), although in terms of absolute aggregate size the distribution at 250 mM is clearly shifted downward.

While *myo*-inositol did not appreciably alter overall aggregate morphology, the more subtle shifts in aggregate size distributions presented a seeming paradox. At 100 mM, the compound reduced the number and size of smaller aggregates but not of larger ones, while at 250 mM it decreased the size of the larger aggregates more than the smaller (indeed, the extended fibril-like aggregates appeared to lengthen). As shown in **Figure 4**, we can explain this behavior by considering avidity effects in protein aggregation. If *myo*-inositol's mode of action involves binding and blocking surfaces involved in the aggregation process, then the lower concentration may only suffice to slow the earliest steps of aggregate assembly, up to the formation of short fibrils, which assemble predominantly one protein-protein interaction at a time. Larger, globular aggregates form predominantly via condensation and coalescence of smaller ones, as evidenced by internal cavities frequently observed within them. Collapse requires formation of many protein-protein interactions simultaneously. High [*myo*-inositol] may inhibit both the early and the late aggregation processes, thus shifting the size distribution back toward the smaller aggregates while further slowing down aggregation overall.

**Microscopic mechanism of aggregation suppression**

To investigate the molecular-level biophysical mechanism of aggregation suppression, we analyzed thermal melts of the wild-type N- and C-terminal domains of HγD by differential scanning fluorometry (DSF) with the SYPRO Orange dye as hydrophobicity probe (**Figure 5**). Both domains showed small but clear and consistent thermostability increases with increasing [*myo*-inositol]. Osmolytes are known to stabilize protein native conformations in many cases, including, in at least one study, *myo*-inositol.[32] However, these effects typically occur at concentrations an order of magnitude above those studied here. That *myo*-inositol increased stability even in the low-mM range further supports its likely physiological relevance to delaying the age of onset of lens turbidity due to aggregation of misfolded crystallins.

The DSF experiments also presented a paradox, however. As shown in **Figure 5A**, no increase in overall hydrophobicity could be detected until after ~50 °C for the N-terminal domain, and even at those higher temperatures, *myo*-inositol's effect was modest (up to ~1 °C of stabilization). Yet, the N-terminal domain aggregated already at 44 °C and was rescued well by *myo*-inositol (**Figure 1D**). The likely explanation is that, as postulated in our prior work, aggregation proceeds from a very early unfolding intermediate, without broad denaturation. The change in hydrophobicity upon formation of this very early intermediate might not be large enough for an external fluorescent sensor such as SYPRO Orange to detect. We therefore carried out thermal scanning Raman spectroscopy to detect the subtle conformational changes that occur in the near-physiological temperature range that is permissive for aggregation.

We placed the protein samples, in solution, in sealed vessels under a Raman microscope to avoid sample evaporation that would otherwise preclude use of this technique for thermal scanning (**Figure 6A**). Temperature was progressively ramped up from 25 to 65 °C to investigate the early unfolding transitions for both the N- and C-terminal HγD domains with and without 100 mM *myo*-inositol. **Figure 6B** shows representative full Raman spectra. At temperatures 35 to 45 °C we observed no major differences in other spectral regions in the presence vs. absence of *myo*-inositol. Specifically, the tryptophan Fermi doublet (1340 and 1360 cm$^{-1}$), skeletal vibration, and phenylalanine pockets (1002-1004 cm$^{-1}$)[33] showed no significant changes.

The amide I region (1600 to 1700cm$^{-1}$) reports mostly (70-85%) the C=O stretch, with a smaller contribution from C-N stretch; it is therefore an excellent probe of protein secondary structure. Absolute intensity of the Raman spectra at 1650 cm$^{-1}$, the typical mid-point of the amide I band, showed a cooperative thermal transition centered at ~45-46 °C for either domain of the protein (**Figure 6C,E**), far below the gross denaturation midpoints of ~63-70 °C by DSF (**Figure 5A**). Thus, Raman spectroscopy revealed a very early unfolding transition not visible to the more coarse-grained DSF technique. For the N-terminal domain only, *myo*-inositol shifted this transition mid-point ~6 °C higher while also reducing its cooperativity – strong evidence of a stabilizing role of *myo*-inositol in the physiologically relevant temperature range.

A major advantage of Raman spectroscopy is the ability to deconvolute Amide I spectra to resolve contributions of specific classes of secondary structure: α-helices, β-sheets, and loop/turn regions. This deconvolution showed the relative contributions of native β-sheet and α-turn structures were unaltered by *myo*-inositol during the observed transition (**Figure 6D,F**). However, at still lower temperatures, 30-40 °C, *myo*-inositol appeared to increase the proportion of native β-structure specifically in the N-terminal domain; α-helical structure remained unchanged, so the proportion of more disordered loop/turn structure became smaller (**Figure 6D**).We observed clear evidence at ~1616 cm$^{-1}$ of non-native, flat β-structure arising at physiological temperatures only in the N-terminal domain, and of *myo*-inositol suppressing this conformational transition (**Figure 6G,H**) until as high as 50 °C (**Figure 6I**).

Raman spectra are extremely information rich and can be used to investigate both global and local structures. Representative full spectra are presented in **Figure 6E**. Quantification of secondary structure contents from the amide I spectra (**Figure 6F**) showed that stabilization by *myo*-inositol did not extend to the helical structures (there are only two helical turns per domain in this protein), while also confirming that.

In summary, thermally scanned Raman spectroscopy allowed us to resolve an early unfolding intermediate characterized by disordering of a fraction (up to ~20%) of the native β-sheet structure. This disordering was specifically mitigated by *myo*-inositol, which did not affect the loss of β-sheet structure in the highly homologous C-terminal domain, nor the loss of the helical structure in the N-terminal domain. Our prior structural model of γ-crystallin aggregation[10,11] entails detachment of the N-terminal β-hairpin with concomitant disordering of just 1 or 2 of the 8 native β-strands in the N-terminal domain. Both the microscopic Raman observations and the TEM imaging of supramolecular morphology are consistent with such a model and point to *myo*-inositol's likely mode of action of inhibiting this local conformational transition.

## **Discussion**

In previous work,[8-11] we have established an *in vitro* model of cataractogenic oxidative aggregation of lens γ-crystallins under physiologically relevant conditions. Importantly, this model reproduces and explains cataract-associated non-native disulfide bonding patterns found by *ex vivo* proteomic analysis of human lenses[34,35] and genetic evidence from cataract-causing point mutant alleles.[5,12] We have now shown that, under the same oxidizing conditions, aggregation is suppressed by *myo*-inositol in a concentration-dependent manner and at physiologically reasonable concentrations. Aggregation suppression by *myo*-inositol is robust to experimental conditions, including the choice of aggregation-permissive HγD variant. Therefore, *myo*-inositol, alongside smaller amounts of *scyllo*-inositol, glucose, and several other sugar alcohols, likely serves as a native chemical chaperone in the human lens, delaying the onset of light-scattering aggregation of lens crystallins. This native chemical chaperone system is depleted during cataractogenesis and may account for some of the variation in the crucial variable for this globally prevalent disease – the age of onset.

Several lines of evidence indicate a direct protein-compound interaction, rather than a generalized through-water osmolyte effect. First, aggregation suppression is already significant in the low-mM range, when [*myo*-inositol] is well below 1% w/v (**Figure 1A**). Second, compositionally similar compounds have highly distinct effects; thus, *myo*-inositol ($C_6H_{12}O_6$) is a much stronger suppressor than mannitol ($C_6H_{14}O_6$), while galactose and IPTG have opposite effects (**Figure 1B**). Third, the stabilization effect is specific to the N-terminal domain (**Figure 6**), despite its very high structural homology to the C-terminal domain.

All aggregation processes observed here proceed at 37-44 °C, well below the global melting transition as seen by DSF (**Figure 5**) and as observed also in prior studies.[8,9] Thermal scanning Raman spectroscopy, however, revealed a sigmoidal thermal transition for either wild-type domain within the physiological temperature range – evidence that a very early unfolding intermediate is the likely aggregation precursor. Accordingly, only a subset of the native β-sheet was lost during the observed early unfolding transition (**Figure 6C,E**). *Myo*-inositol delayed this transition and reduced its steepness (**Figure 6C**) by stabilizing native β-sheet structure at low temperatures (**Figure 6D**) and inhibiting conformational transition to non-native β-sheet at higher temperatures (**Figure 6G,H,I**).

Unlike amyloid deposition diseases, EM imaging of cataractous lenses revealed not the characteristic long straight fibrils but only an increased texturing of lens cytosol, sufficient to cause the light scattering that is the hallmark of this disease.[36,37] Our present observations in **Figure 2** and other studies *in vitro* have likewise supported a predominantly non-amyloid character of crystallin aggregates under the most physiologically relevant conditions.[8,9,38,39] However, infrared spectroscopy did reveal significant flat β-sheet content in the cataractous lens,[40] and our Raman data on the aggregation-prone N-terminal domain concur. Local regions of flat β-sheet structure within the context of topologically rearranged domain swap-like aggregates are consistent with the hairpin-swapping model we have previously proposed[10] for these natively β-sheet-rich proteins.

Additional research is needed to determine whether *myo*-inositol stabilizes other lens crystallins and whether other compounds based on the inositol scaffold are even better aggregation suppressors. For example, since redox chemistry is crucial in crystallin aggregation[10,11] and glutathione fails to diffuse into the aging lens,[41] replacing one hydroxyl of *myo*-inositol with a thiol to might yield a bifunctional anti-cataract drug ("thio-inositol") – a structural stabilizer and a reducing agent. Free inositol concentrations need to be measured in healthy lenses from across the lifespan, as existing datasets have focused on aged cataractous lenses and lenses from age-matched controls. The observation that young lenses have a lower free water content[30] would suggest that effective [*myo*-inositol] is highest in young age. The structural determinants of crystallin stabilization – including binding site(s) – remain to be determined. Finally, whether and how this chemical chaperone synergizes with the proteinaceous chaperones (αA- and αB-crystallin) should be a subject of future research. Lens inositol depletion in a mouse knock-in of a human cataract-associated αB-crystallin variant was recently reported.[42]

A typical human diet is estimated to contain 1-2 g of *myo*-inositol per day, and ~4 g/day is produced endogenously from glucose in the liver and kidney.[43] Of all human tissues in which *myo*-inositol levels have been studied, the lens appears to have the highest, while the second-highest levels are found in the cytoplasm of nerve cells in the brain.[43] High brain inositol has been ascribed to the need for signaling processes via second-messenger cascades,[43] but it may also act as a chemical chaperone there. Thus, both *myo*- and *scyllo*-inositol, as well as some derivatives, have been shown to directly bind and stabilize oligomers of Aβ42, inhibiting amyloid aggregation.[44,45] *Scyllo*-inositol has been clinically studied in Alzheimer's disease in humans, with some promise as a disease-modifying agent.[46] Our finding of *myo*-inositol's likely aggregation-suppression role in the lens calls for more detailed investigation of whether such activity is general for β-sheet rich proteins.

Several animal and human studies have observed a relationship between *myo*-inositol levels and cataract. Thus, in rats with streptozotocin- or galactose-induced cataracts, dietary supplementation with *myo*-inositol dramatically decreased the rate of cataract development: by 14 weeks of age, the untreated animals had totally opaque lenses, while the treated showed only initial stages of cataractogenesis.[47] In another study, dietary supplementation with *myo*-inositol triphosphate delayed cataract onset by ~44% in streptozotocin-induced diabetic rats.[48] The premise of the rat studies was that *myo*-inositol might act as an inhibitor of aldose reductase in the lens, but this was found not to be the case. Direct inhibition of protein aggregation was not considered at the time. Dosage of *myo*-inositol is important, however, as genetic overexpression of its native lens transporter in mice resulted in cataract due to hyperosmotic stress.[49] Accordingly, trisomy-21 (Down's syndrome), which leads to overexpression of the same transporter in humans, is likewise associated with premature cataract.[50] Given that *myo*-inositol is dramatically depleted in cataractous human lenses but not in age-matched controls,[26,27] and given that it is one of the very few molecules able to diffuse into lens tissue, maintaining or restoring healthy levels of lens *myo*-inositol should be considered a promising strategy for delaying cataract onset and thus reducing the global burden of this disease.

## **Materials and Methods**

### **Protein expression and purification**

Wild-type and mutant human γD crystallin, without any tags, was overexpressed in BL21 RIL *E. coli* from a pET16b plasmid and purified by ammonium sulfate fractionation, ion exchange, and size exclusion as described previously.[11]

### **Aggregation assays (turbidimetry)**

Aggregation was induced by heating at 37 °C (unless otherwise indicated) in the presence of 0.5 mM oxidized glutathione as described previously.[11] Good's buffers were added to a final 10 mM concentration to set sample pH to a desired level: MES for pH 6.0, PIPES for pH 6.7, and HEPES for pH 7.4. Experiments in the main text were conducted in PIPES pH 6.7 buffer, unless otherwise indicated, because intralenticular pH is known to be slightly below neutral.[51] All compounds tested as aggregation inhibitors were of high purity. *Myo*-inositol (>99%), *scyllo*-inositol (>98%), glucose (>99%), sucrose (>99.5%), trehalose (>99%), and rhamnose (>99%) were purchased from MilliporeSigma (Burlington, MA). Galactose (98%) and arabinose (99%) were from Alfa Aesar (Heysham, UK); mannitol (99%) from BeanTown Chemical (Hudson, NH); glycerol (99.7%) from Avantor (Radnor, PA); and IPTG (99%) from Omega Scientific (Tarzana, CA).

### **Differential scanning fluorometry**

Samples were buffer exchanged into 10 mM PIPES pH 6.7 buffer with no added salt (estimated 15 mM [Na$^+$] final) by gel filtration and used at 40 µM for the thermal melts. Lack of salt was found empirically to minimize aggregation. A reducing agent (1 mM tricarboxyethyl phosphine (TCEP)) was added to prevent oxidative misfolding. 1X SYPRO Orange (Thermo Fisher) was added as the hydrophobicity probe. Melts were carried out using Bio-Rad CFX 96 Touch thermocyclers, with temperature ramping of 1 °C per minute between 25 and 95 °C. Samples were prepared by mixing a 2X protein sample with the appropriate ratio of aqueous *myo*-inositol and water and a constant amount of PIPES buffer to avoid any variation in pH or salinity among samples.

### **Transmission Electron Microscopy**

Samples were prepared in 10 mM PIPES pH 6.7 with 150 mM NaCl with 1 mM EDTA and 0.5 mM oxidized glutathione (except the control sample, which lacked glutathione) and incubated at 37 °C in the presence or absence of *myo*-inositol as indicated for 4 h. They were then kept at room temperature for ~2 h before being applied to carbon-coated 200-hex copper grids (EMS, Hatfield PA) prepared following the manufacturer's instructions and negatively stained with 2% uranyl acetate. Imaging was carried out on a Hitachi 7800 TEM system. For maximal precision, aggregate classification and morphometric measurements were carried out manually using contour and area tracing in ImageJ on a stylus-equipped touchscreen. To minimize human bias during this morphometric analysis, a double-blind was implemented as described in Results.

### Raman Spectroscopy

To better understand the impact of inositol on crystallin structure we went on to carry out Raman spectroscopy. Raman spectroscopy as a vibrational spectroscopy technique provides a host of structural information and has been a popular choice for spectroscopy of protein molecules[52]. Compared to other spectroscopic techniques, Raman spectroscopy is essentially a non-destructive technique and has a very high sensitivity. Raman has been specially an attractive technique to probe into the secondary structural aspects of a protein molecule[53]. Conformational changes which primarily impacts the secondary structure of a protein can be potentially captured using amide I spectral components[54]. Compared to FT-IR spectroscopy, one single biggest advantage in Raman spectroscopy is that it is largely devoid of spectral interference from water in the amide I range. Further additional information rich components like Tryptophan Fermi doublets and phenylalanine pockets make Raman a useful technique in probing global protein structure and probe under conditions of perturbation[55]. In our study we aimed to investigate how the N and C terminal domains of Crystallin essentially gets impacted upon thermal perturbation. Further we wanted to probe how myo-inositol impacts the thermal transition trend for both the domains. It is to be noted that performing thermal perturbation with in-solution proteins is a challenging task. The process tends to suffer from evaporation issues as we keep ramping up the temperature leading to anomalous in-solution conditions. To address this issue we developed a novel apparatus to allow thermal scanning while preventing evaporation and minimizing sample handling.

### The thermal scanning Raman spectrometer

Raman spectroscopy was carried out using Horiba XploRa confocal Raman microscope, a detector thermoelectrically cooled to -70 C and a 1200 gr/mm grating blazed at 750nm. For these experiments, a solid-state 785 nm laser was used for excitation. An acquisition time of 180 seconds was used for all measurements. Two back-to-back spectra were collected to allow for automatic removal of cosmic rays. The Horiba denoise algorithm in standard mode was used to slightly smooth the spectra. The fluorescent baseline was fit and subtracted using a polynomial fit before peak deconvolution was performed in Labspec 6. Power levels at the sample for the 785 nm laser excitation typically remained around ~41 mW. The spectral resolution was 1 $cm^{-1}$. As the excitation wavelength is distant from any protein absorption bands, photo-bleaching or heat induced sample deterioration was not expected. The spectrometer slit was set to 200 µm. The confocal aperture was set to 500 µm. The system was calibrated to the 520.7 $cm^{-1}$ silicon reference sample before every set of measurements. 20 µM protein in 10 mM PIPES buffer pH 6.7 was used for the measurements with a sample volume of 20µl. Samples with inositol had 100mM inositol. Since the Raman microscope was confocal, the measurements could be made through a sealed PCR tube with no sample handling. The PCR tube was laid on its side on a microscope slide with a slight tilt to keep the protein solution in the bottom of the tube. The Raman microscope was then focused into the middle of the protein solution at the bottom of the tube. The thermal scanning was done on a home-built temperature stage by taping a 10 kΩ 25 W resistor (Digikey MP825-10.0K-1%-ND) with Kapton tape to the microscope slide. A thermocouple was placed between the PCR tube and the resistor to monitor sample temperature. DC voltage was applied to the resistor to heat the sample in a controlled manner. By slowly increasing the voltage and monitoring the thermocouple readout, the temperature could be increased in a controlled manner. The PCR tube was secured on top of the resistor with Kapton tape. Kapton tape was essential for the setup to be able to reach 65 °C. (See **Figure 6A** for schematic of thermal stage setup.) Thermal scans were done on the same sample progressively heating the sample in 5 °C steps. Buffer blanks were carefully taken with buffer and buffer + inositol for protein and protein + inositol sets. Spectral readouts were normalized with respect to the sharp peak at 330 cm$^{-1}$.

To ensure the quality of the protein specific signal, we recorded spectra of WT and compared that with PDB 1HK0. A comparison of the secondary structure contents as retrieved from our Raman Amide I spectra showed comparable contents with 1HK0.

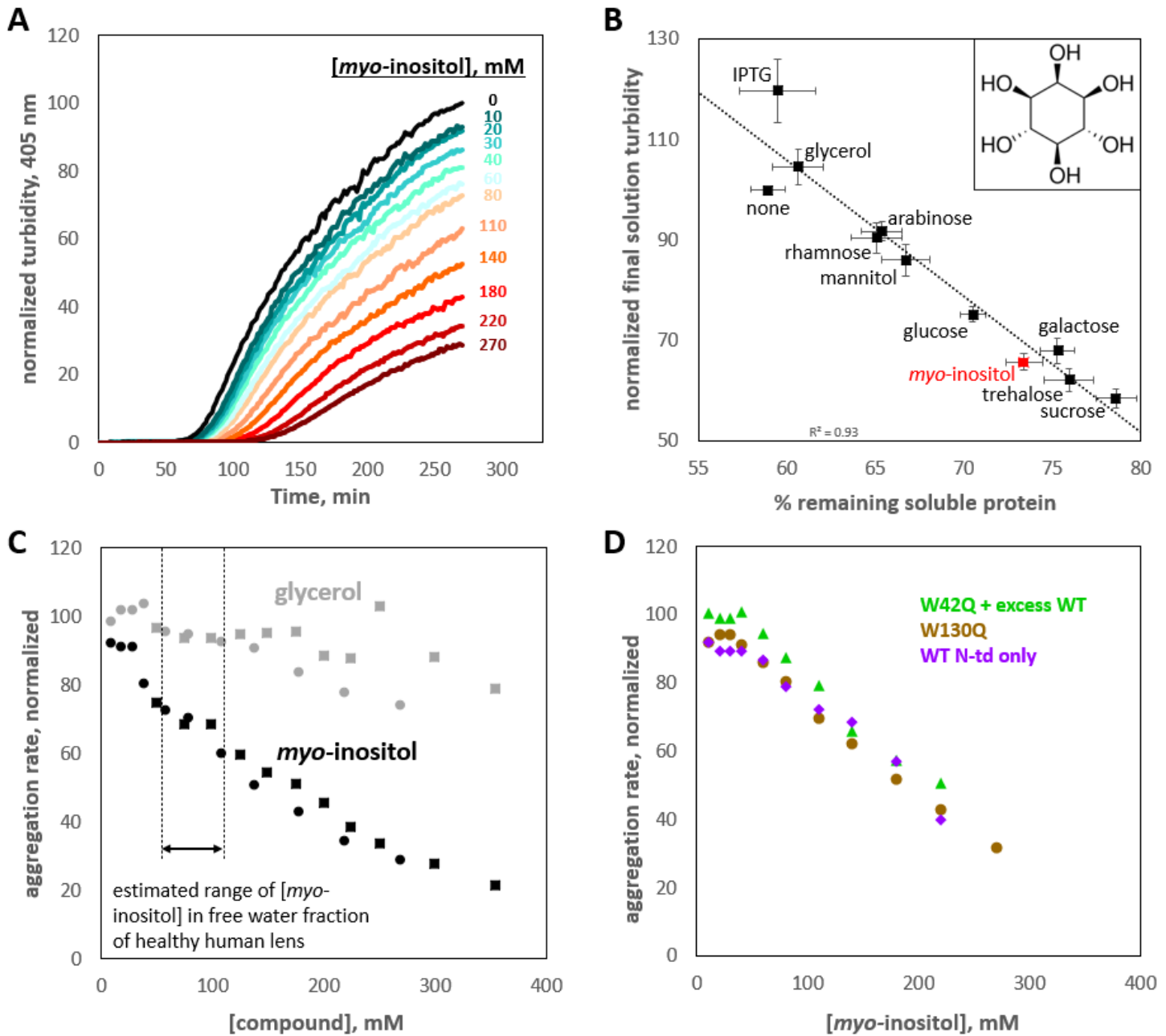

**Figure 1: Suppression of human γD-crystallin aggregation by *myo*-inositol.** Oxidative aggregation of the cataract-mimicking W42Q variant of HγD was initiated as previously described.[11] (A) Normalized turbidity traces for the oxidative aggregation of 40 μM HγD W42Q with varying concentrations of *myo*-inositol. (B) Small carbohydrates, each at 100 mM, suppressed HγD W42Q aggregation to varying degrees. Isopropyl-β-D-thiogalactoside (IPTG) moderately enhanced turbidity development. *Myo*-inositol (structure shown in *inset*) consistently and strongly suppressed turbidity development, second only to the disaccharides, trehalose and sucrose. A total of 8 independent replicates, at pH 7.4 (HEPES), pH 6.7 (PIPES), or pH 6.0 (MES) and 150 mM NaCl all produced similar results and were averaged together. Notably, the strong linear correlation between reduction in solution turbidity and increase in the proportion of protein remaining soluble indicated these compounds (except perhaps IPTG) did not significantly change aggregate geometry. (C) Dose response for aggregation suppression by *myo*-inositol compared to glycerol. Notably, *myo*-inositol had a significant effect in the physiological concentration range. Data from two independent replicates (circles, squares) are shown; *black circles* correspond to the data in panel A. All aggregation rates were normalized to the rate without *myo*-inositol. (D) Suppression of oxidative aggregation by *myo*-inositol generalizes to other HγD constructs: G*reen triangles*, 20 μM W42Q whose aggregation is catalyzed by 180 μM WT HγD at 37 °C; *Beige circles*, 40 μM W130Q at 42 °C; *Purple diamonds*, 50 μM wild-type isolated N-terminal domain of HγD at 44 °C. All experiments were carried out in 10 mM PIPES buffer pH 6.7 with 150 mM NaCl, 1 mM EDTA, and 0.5 mM GSSG as the oxidant.

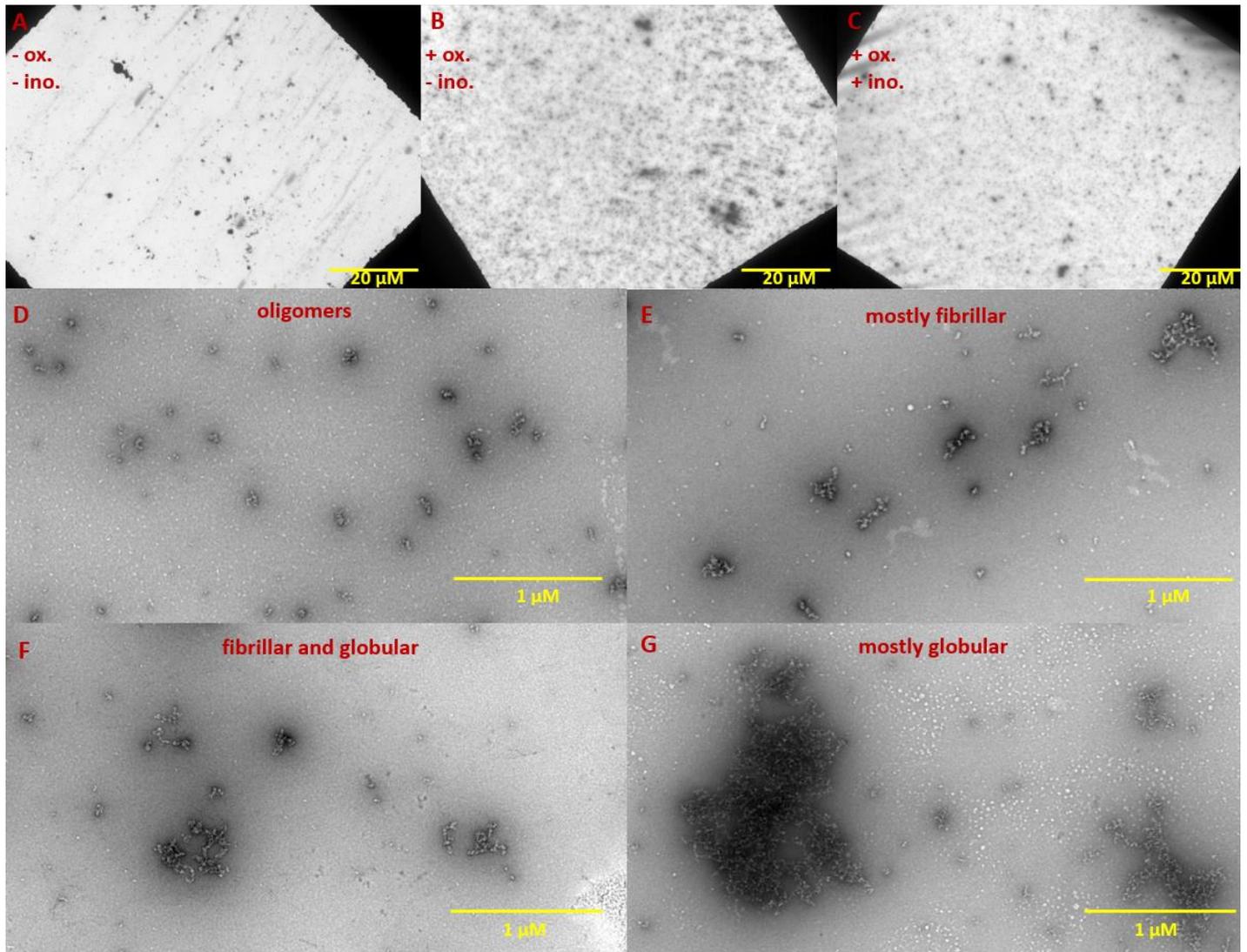

**Figure 2: Negative-stain TEM images of HγD W42Q aggregates with and without *myo*-inositol.** Aggregation of 40 µM protein in pH 6.7 PIPES buffer with 150 mM NaCl and 1 mM EDTA was triggered by addition of oxidant (0.5 mM GSSG) and incubation at 37 °C for 4 h. Samples from the end-point of the assay were deposited on carbon-coated copper grids and stained with uranyl acetate. The top row shows survey images of whole grids for (A) control sample incubated in the absence of GSSG; (B) turbid sample in the presence of GSSG; and (C) turbid sample in the presence of GSSG and 100 mM *myo*-inositol. Panels D-G show the variety of aggregate morphologies observed in representative magnified images from these grids, with (D) showing mostly small oligomers (not counted as aggregates in **Figure 3**); (E) showing mostly fibrillar aggregates; (F) showing fibrillar aggregates in the process of collapsing and coalescing; and (G) showing highly coalesced globular aggregates.

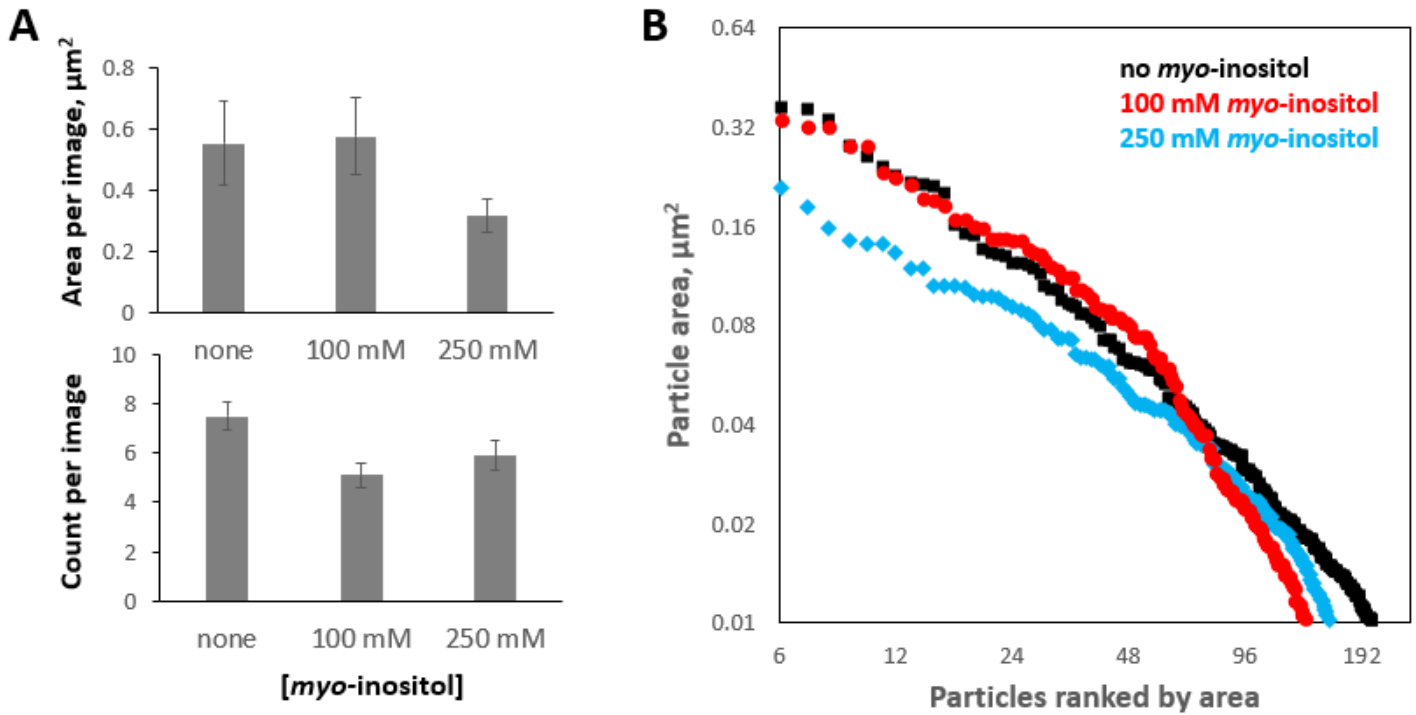
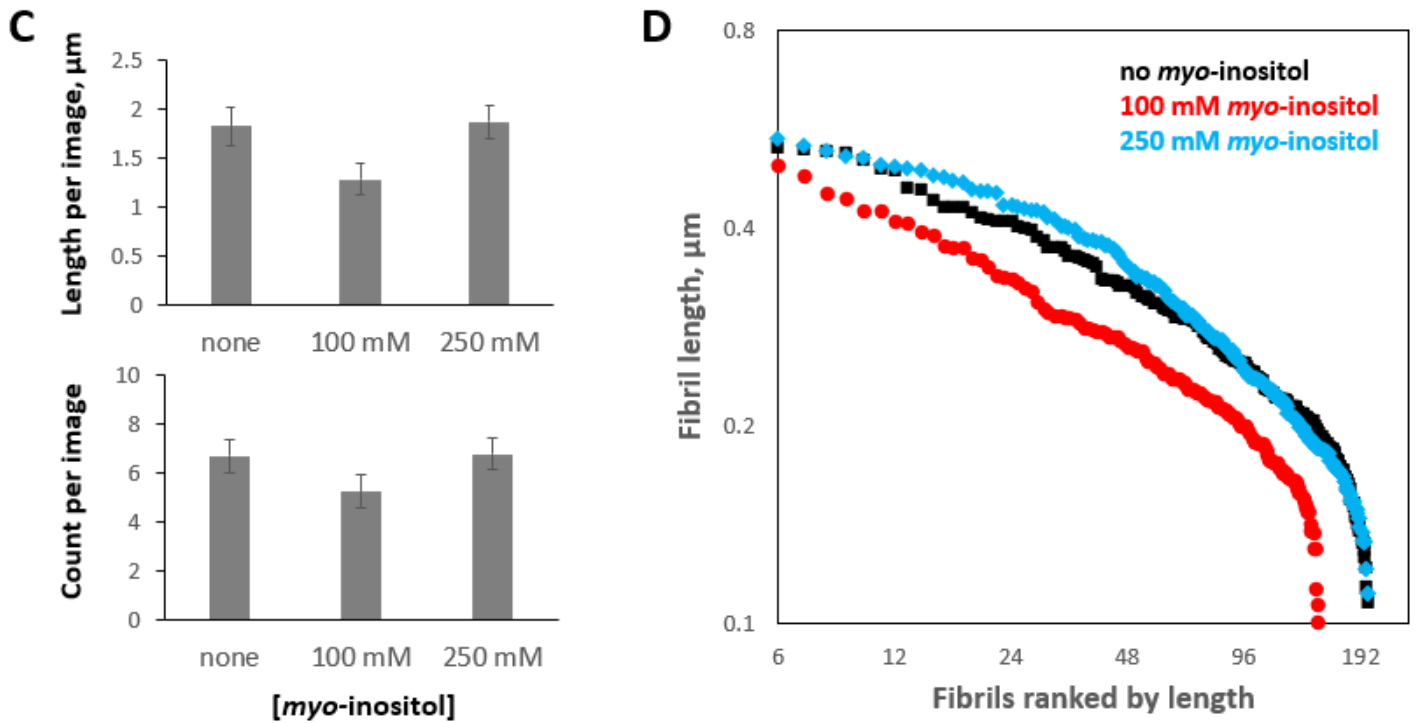

**Figure 3: Tell-tale shifts in aggregate size distribution in the presence of *myo*-inositol.** Aggregates generated as in Figure 1 in the presence of 0, intermediate (100 mM) or high (250 mM) [inositol] were imaged and analyzed by negative-stain transmission electron microscopy in a double-blinded fashion, with a total of 103 separate images used for analysis. Aggregates, which were composed of highly flexible branching fibrils, were classified by inspection into globular or fibrillar based on whether a curve could be clearly traced from end to end. (A, B) 100 mM *myo*-inositol selectively depleted smaller globular aggregates, while 250 mM depleted larger ones. (C, D) By contrast, fibrillar aggregates (too small to have undergone collapse to globules) became slightly longer in 250 mM inositol but were depleted at 100 mM across the board.

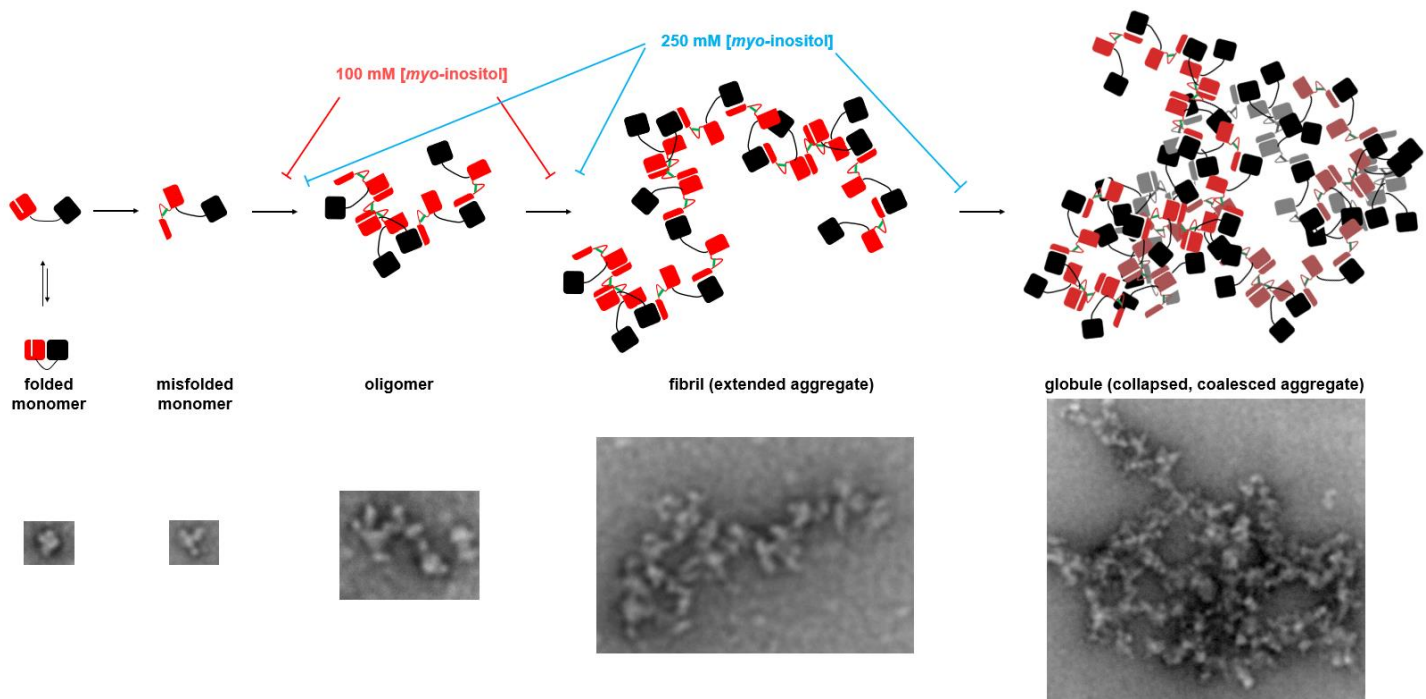

**Figure 4: Graphical model of HγD aggregation and its suppression by *myo*-inositol.** The aggregation process begins with oxidative misfolding of a mutant or damaged monomer molecule, as we have previously determined.[10] Next, such molecules assemble via domain swap-like interactions, as proposed earlier.[10,11] Examples of TEM evidence in this study (*bottom row*) are consistent with growth of short extended structures (fibrils) by the proposed mechanism, followed by coalescence and collapse of the small aggregates to form larger particles of a roughly globular shape. We hypothesize that the differences in size distributions observed in Figure 3 are explained by avidity effects during the aggregation process. Early-stage aggregation occurs via addition of monomers or small oligomers, interacting at one or more specific binding surfaces, which is inhibited by [*myo*-inositol]. Later-stage collapse and coalescence, however, are driven by simultaneous interactions of a multitude of specific and non-specific interactions, which only the highest [*myo*-inositol] can partially inhibit.

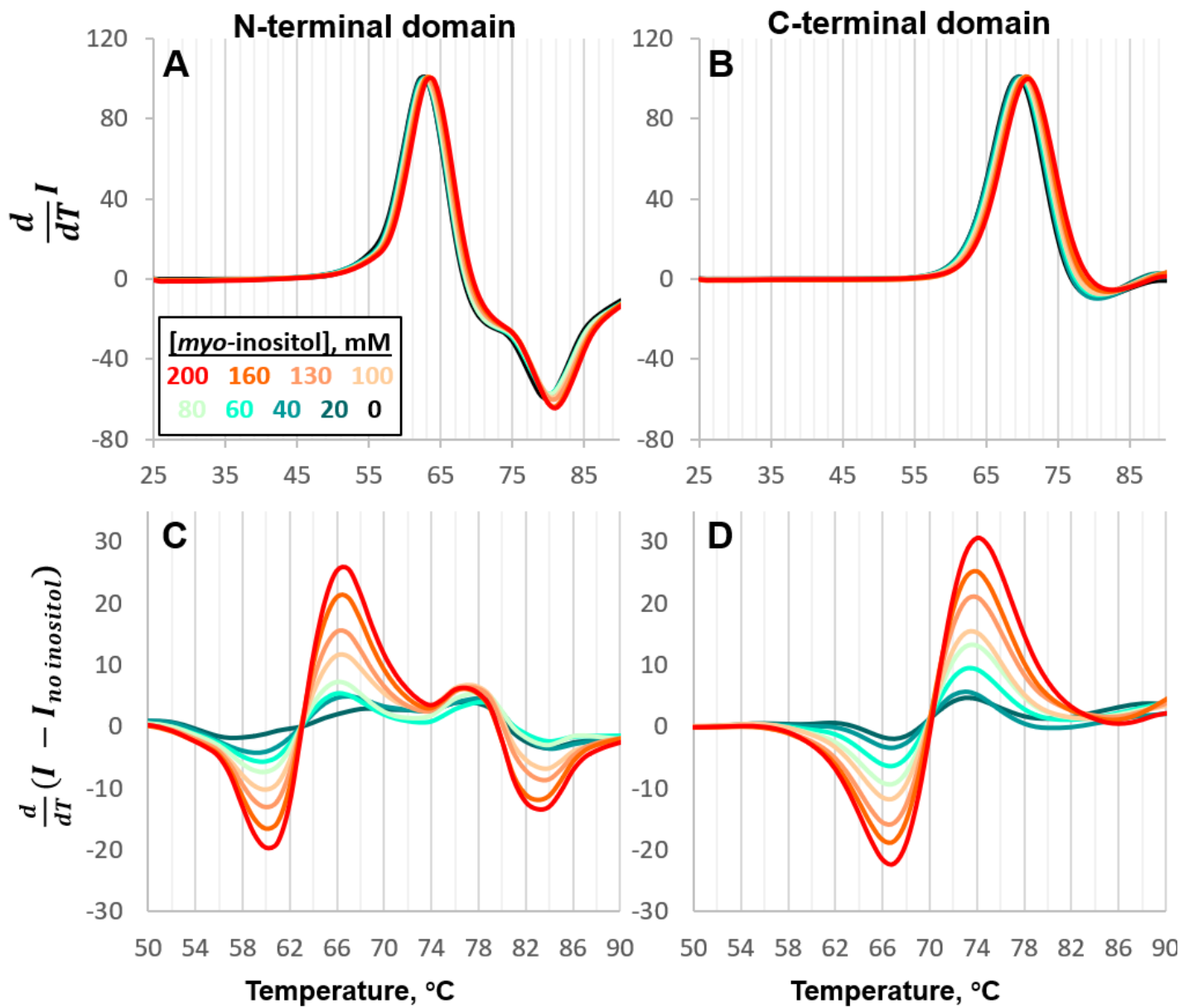

**Figure 5: *Myo*-inositol stabilizes both domains of HγD.** Differential scanning fluorometry with SYPRO Orange as the hydrophobicity probe revealed small but clear shifts toward higher melting temperatures with increasing concentrations of the compound. (A, B) Normalized first derivative of the dye fluorescence intensity; peak positions correspond to $T_m$. The N-terminal domain exhibits both a positive peak (transition to molten globule that binds the dye) and a negative peak (transition to a fully unfolded state no longer containing hydrophobic pockets needed for dye binding). (C, D) Subtracting out the no-inositol traces in (A, B) shows clear and monotonic build-up of residual intensities with increasing [*myo*-inositol], starting from the lowest (sub-physiological) concentrations of the compound used in these experiments.

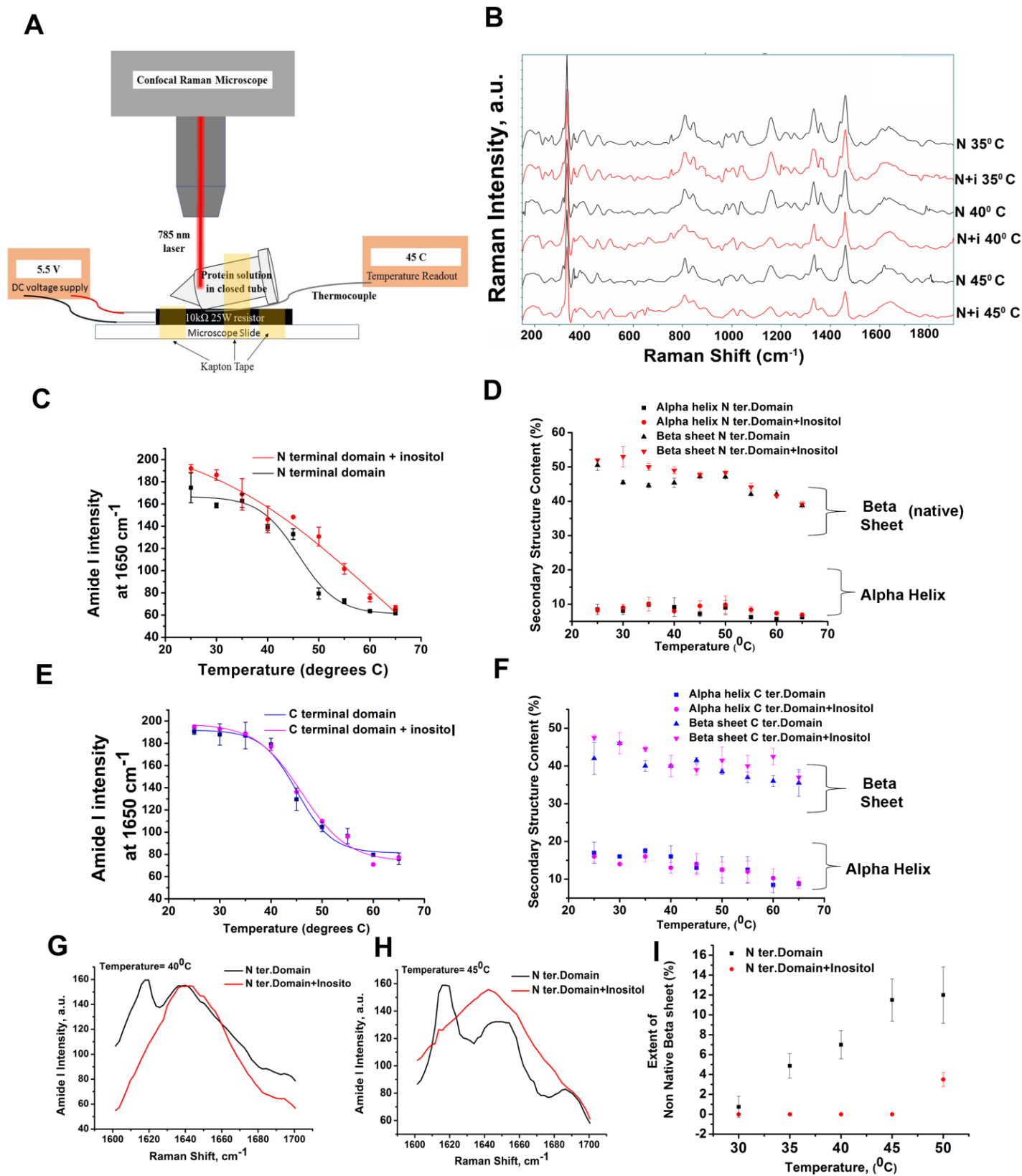

**Figure 6: Thermal scanning Raman spectroscopy reveals effect of *myo*-inositol on local secondary structure of the N terminal domain.** (A) Schematic depiction of the Raman spectroscopy set-up deployed for the thermal perturbation study (details in *Materials and Methods*). (B) Raman profiles of the N terminal domain in presence (*red*) and absence (*black*) of *myo*-inositol showing that overall structure of the protein is preserved. (C) Intensities of Raman spectra for the N-terminal domain with and without *myo*-inositol from 25 °C to 65 °C measured at 1650 cm$^{-1}$, the typical midpoint of the Amide I band,

which reports on protein secondary structure content. Empirical sigmoid fits showed a less cooperative transition and a shift in transition mid-point from 46 °C to 52 °C in the presence of *myo*-inositol. (D) Relative proportions of secondary structures in the N terminal domain upon thermal perturbation, in presence and absence of inositol, not including non-native β-structure. Amide I spectra were deconvoluted using Lorentzian fit and peaks assigned using standards for specific secondary structure types. Non-native extended β-structure was included in the fitting but excluded from this plot to focus on the native secondary structure. Error bars show S.E.M. of 3 biological replicates with 2 technical replicates each. (E,F) Same as (C,D) but for the C terminal domain. Addition of inositol did not measurably impact this domain's Raman spectra in this temperature range. (G) Amide I spectra of N terminal domain (mean of 4 independent acquisitions) at 40 °C clearly showed a peak at ~1616 cm$^{-1}$ which is a characteristic of non-native β-sheet and has been traditionally attributed to protein aggregation (*black*). No such peak was observed in presence of *myo*-inositol (*red*). (H) Same as (G) but at 45 °C, showing a decline in overall intensity of the native structural content as the aggregate peak begins to dominate the spectrum. *Myo*-inositol prevented this decline and inhibited formation of non-native structure. (I) Non-native β-sheets in the N-terminal domain formed at much lower temperatures in the absence of *myo*-inositol.

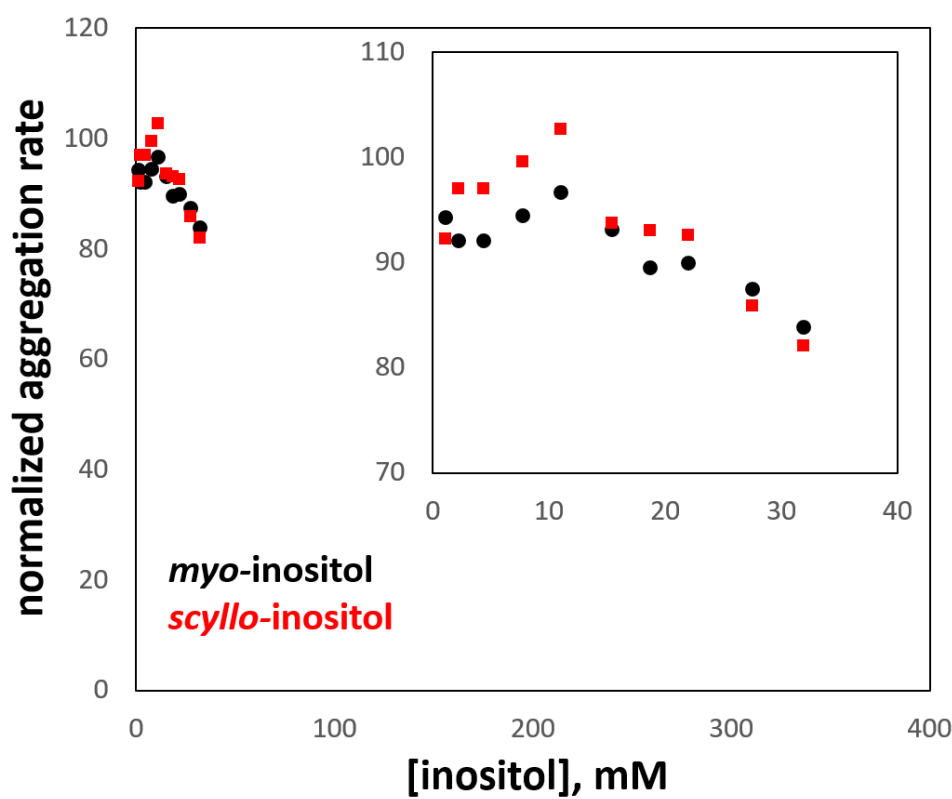

**Figure S1: Comparison of *myo*- and *scyllo*-inositol.** Both compounds were tested in parallel under the same conditions for their ability to suppress aggregation of the W42Q HγD variant, exactly as in Figure 1. However, *scyllo*-inositol solubility is an order of magnitude lower than that of the more common *myo*- isomer, which strictly constrained the testable concentration range. In this range, no significant difference was observed between *myo*- and *scyllo*-inositol. The axes are scaled to match Figure 1C,D; the inset shows a zoom-in of the data series. All aggregation rates are expressed as % of the rate without inositol.